\documentstyle[11pt]{article}
\begin{document}
\begin{center}
{\Large It's a Wrap! Reviewing the 1997 Outdoor Season} \\
\vspace{5mm}
J. R. Mureika \\
{\it Department of Computer Science \\
University of Southern California \\
Los Angeles, CA~~90089-2520}
\end{center}

Throughout the summer, I've written articles highlighting this year's
100m performances, and ranking them according to their wind-corrected
values.  Now that the fall months draw to a close, and the temperature 
drops to a nippy 15 celsius at night (well, for some of us), it seems
only natural to wrap up the year with a rundown of the 1997
rankings. \\

Of course, it wouldn't be exciting to just give the official rankings, 
so I will also present the wind-corrected rankings, and will offer comparison
to the adjusted value of the athlete's best 100m performance of 1996.
Mind you, this won't necessarily be the {\it best} wind-corrected performance,
but it can offer an insight into how an athlete has progressed over the
course of a year. \\

 As a quick refresher, a 100m time $t_w$ assisted by a 
wind $w$ (the wind speed) can be corrected to an equivalent time $t_0$ as 
run with no wind ($w=0\;$m/s), 

\begin{equation}
t_0 \approx \left[ 1.03 - 0.03\times \left(1 - \frac{w \times t_w}{100}\right)^2
\right]\times t_w~.
\label{wind}
\end{equation}

\noindent This comes about because approximately 3\% of the athlete's 
effort is spent
fighting atmospheric drag.  A tail-wind implicitly boosts a race time,
hile a head-wind can take away a sprinter's chance at a possible World Record.
(as an interesting aside: this also says that it takes more energy to run 
into a head-wind than you get from a tail-wind assistance. So, a 400m can be
handicapped by a sufficiently strong wind, since running
with the wind on one straight does not compensate for fighting the head-wind
on the other straight). \\

  The next logical step in this review would be to make some
predictions for the upcoming indoor season, based on how everyone 
performed outdoors over 50 and 60m (using appropriate splits from Athens,
and other meets when available).  Tables~\ref{table1},~\ref{table3} 
show the top 5 wind-corrected rankings for men and women, while 
Tables~\ref{table2}, \ref{table4} give the official rankings.  \\

\noindent{\bf Men's 1997 Rankings} \\

This summer's competition was a veritable roller coaster for both the men
and women, full
of very unexpected ups and downs.  One of the bigger surprises was
the rise of Maurice Greene to the top-ranked sprinter of the year.  His 9.86s
victory (9.88s wind-corrected) marked almost a 0.2s improvement over his
1996 PB of 10.08s (with -0.3 m/s wind; 10.07s wind-corrected), which was
ranked 28th in the 1996 IAAF list!  Of the top 5 wind-corrected times,
his showed the largest improvement of 1.9\%.  Bailey came close to matching
his best from '96 (that is, his wind-corrected PB) with his Abbotsford run
of 9.89s  (10.03s, -2.1 m/s).  Ato Boldon chimes in at 3rd fastest after
wind-correction, and although he clocked the most sub-9.90s performances
of the year (9.89s in Modesto, and 9.87s in Athens), his best wind-corrected
performance came out of Stuttgart.  Since this time was run in still
conditions (9.90s, +0.0 m/s), it's an {\it official} wind-corrected time!
However, it also matches his best wind-corrected performance of
1996 (9.90s; 9.93s, -0.6 m/s).    \\

Fredericks rounds out fourth on the 1997 rankings, with his late-season
9.92s (9.90s, +0.2 m/s) from Brussels (since he finished 4th in Athens,
it seems that Frank truly was the 4th best in the world this year!).  
Unfortunately, Fredericks also shows the worst improvement from last year
of the five, not being able to match his incredible 1996 wind-adjusted
mark of 9.84s (9.86s, -0.4 m/s) from Lausanne.  Interestingly enough,
5th ranked Tim Montgomery showed similar improvement to Greene since
last year, where he also ran 10.07s (10.08, -0.3).  In fact, Montgomery
was ranked 27th, just above Greene on the 1996 IAAF top 50 list.
A most impressive showing by the Americans in Athens. \\

Although the peak performances of Bailey, Boldon, and Fredericks did not
surpass their 1996 bests, Table~\ref{table2a} presents a quick glimpse of
their top 5 times of each year.  Also present is the average of each time,
and the although not necessarily appropriate for an ordered set of data,
the standard deviation is offered to give an idea of the variance of the
times. Only Fredericks does not better his 1996 average,
while Bailey tops his by 0.02s, and Boldon by 0.01s.  It's interesting
to note that both Fredericks and Boldon had faster averages than Bailey
in 1996.  Sometimes, it all comes down to just one race. \\

While Fredericks posted the fastest average of 1996, Boldon takes it for
1997, and although he could not better his top time of 9.90s, he shows
an impressive ability to run repeat fast times over the two years (standard 
deviations of 0.023 asnd 0.029s respectively).  Meanwhile, Bailey has also
improved his knack to do same, but Fredericks clearly takes the cake with
a '97 standard deviation of 0.019s!  One could say that he has consistently
run the most fast times this year.   His top 5 marks for
1997 fall in an interval of 0.05s, compared to Bailey and Boldon's
0.09s interval. \\

As a note of national interest, sixth place is given in Table~\ref{table1}.
Thanks in part to a strong head-wind, the National Championships in Abbotsford 
not only produced Bailey's 2nd place 9.89s, but also gave Robert Esmie a
wind-corrected sub-10s clocking at 9.96s, good enough to rank him 6th in
the world!  His 9.96s clocking shows a 3\% (0.3s!) improvement over his
1996 PB of 10.27A (10.18A, +1.5), which essentially matches his next fastest
time of 10.28s (real still air PB, +0.0 m/s wind).  Experimental evidence
for wind-correction! \\

\noindent{\bf Women's 1997 Rankings} \\

The women's list are no less impressive than the men's.  In fact, they
are in a sense {\it more} impressive.  A quick examination of Table~\ref{table3}
shows that two of the five individuals were not ranked in the top 50 last
year.  Additionally, 3 of the official top 5 in Table~\ref{table4} were
similarly unranked!  The emergence of Marion Jones onto the sprint scene
this year showed a virtually unparalleled rise to the top of the 100m 
podium.  To be a newcomer to the sport and capture gold at the World
Championships in the same year is a feat until itself.  Table 4 attests
to her exclusive domination of the event this summer. \\

On the way up, Jones bumped Jamaican sprint legend Merlene Ottey
down a notch from her 1996 throne.  To the very end, Jones also 
dominated the wind-corrected rankings, but in an impressive show in Tokyo
on September 6th, Ottey chalked up a 10.81s into a 0.4 m/s head-wind.
This adjusted to a 10.79s in still conditions, and put her 0.03s under Jones'
fastest wind-corrected dash (10.82; 10.76s, +0.9 m/s).  It was also her
fastest wind-corrected time ever, bettering her previous adjusted PB of
10.80s, a mark she set twice (Zurich, 07 Aug 1991; and 16 Aug 1993).  Almost
two decades in the sport has certainly paid off for Merlene! \\

Meanwhile, 200m
world champion Zhanna Pintussevich put in a 10.88s (10.85s, +0.4 m/s) showing
to capture 3rd on both the wind-corrected and official lists.  Like Boldon,
Devers could not manage to surpass her 1996 wind-corrected mark of 10.92s
(10.89, +0.5 m/s), well off her  best 10.75s (10.82, -1.0 m/s) from 
the Barcelona Olympics in 1992.   \\

As with the men's list, fifth place is
a variable: Sevatheda Fynes holds the wind-corrected spot with a head-wind
10.98s (11.07s, -1.2 m/s), but officially Christine Arron (also unranked
in 1996!) takes it with 11.03s (-0.3 m/s).  Once again, Christine Arron was
not ranked in the IAAF's top 50 list from 1996!  Each athlete who was
ranked on the 1996 list (Ottey, Devers) showed drops in their top 
performances of 1997. \\

\noindent{\bf Split Comparison to Indoor Marks} \\

Indoor races, by their nature, are not influenced by wind effects for
obvious reasons.  So, most indoor performances should be comparable to
each other on an implicitly equal footing.  The only times which cannot
be immediately compared to others are those sprints which are run at
altitude.  For example, both Donovan Bailey's 50m WR (5.56s) and Obadele
Thompson's 55m WB (``World Best'') of 5.99s were run at altitude, a factor 
which mysteriously increases sprint potential. \\

Since we can correct outdoor races to get an idea of what the times might
have been in still air conditions, it should also be possible to break
down the 100m races to determine the 50m and 60m splits devoid of wind
effects.  To do so, the formula given above requires slight modification:
the 100 in the denominator of the fraction represents the distance of
the race.  So, by replacing 100 with either 50 or 60, we can wind-correct
the splits. \\

Table~\ref{table5} shows the corrected splits for the men's
and women's 100m finals from Athens, which can be compared to the times in
Table~\ref{table6}.  In the case of the men, every 
competitor considered in Table~\ref{table5} marked a faster performance than
the best indoor 50m of last winter (5.60s...  Aside: is this
the first time the men's short sprint lists have been headed by two people
named M. Green[e]?).  Likewise with the 60m splits, the leaders from Athens
surpass the indoor 1997 marks over the same distance.  Ato Boldon shaved
0.03s off his indoor season best (which of course does not account for reaction
time, so technically he may have actually run the same raw time with a 
different reaction). \\
  
Note that in running his Athens final, Maurice Greene tied {\it both} the 50m
and 60m indoor world records for the distances (60m - 6.41s, Andre Cason,
Madrid, 14 Feb 1992). Greene's best indoor 60m mark for 1997 was 6.54s, 
slower than all of the others' top indoor performances. 
At the USATF Championships 100m final in Indianapolis (13 Jun), 
he and Tim Montgomery both split at 50m in 5.57s (5.56s, +0.2 m/s), 0.01s off
the indoor WR.  Although outdoor racing
seems to be, in general, faster than indoor racing (perhaps a physiological
effect?), these findings suggest that interesting things may be in store
for the coming 1998 indoor season! \\

The women's lists are lacking somewhat, due to insufficient data.  Note that
World Bronze medallist Fynes is not listed in Table~\ref{table5} (not my fault;
the appropriate data wasn't recorded).  However, there's enough given to make
the same stab as with the men's lists.  In this case, the splits correct
to times which are much closer to the indoor marks from last winter.  Only
Jones and Pintussevich surpass the 50 leader (6.05s, Privalova), but
Pintussevich betters her indoor mark by 0.2s!  The performances of Arron,
Miller, Paschke, and Ottey are similar to the indoor 50m list.  All times
are slower than Privalova's WR of 5.96s (Madrid, 9 Feb 1995).  Jones
marked a 50m split of 6.13s (6.08, +1.1) at the USATF Championships. \\

For the 60m, Pintussevich again betters her indoor time over the equivalent
distance by roughly 0.2s, but Paschke and Ottey failed to improve theirs.  
In their defense, Paschke's wind-corrected split of 7.18s is only 0.01s
slower than her '97 best indoor (could be a factor of reaction time), while
Ottey was by no means running her perfect race in Athens!  Perhaps someone
should have timed her 60m false start for comparsion? \\

\noindent{\bf{The Shape of Things to Come?}} \\

And so, as 1998 approaches, we gear up for the new indoor season.
Come March, it will be interesting to compare the 1998 indoor lists with
the analysis in this article.  Many times, the 50m and 60m rankings can be
filled with names unknown to the top 10 outdoor performances.  Will
we see such a phenomenon again?  \\

The trend in particle physics is to make predictions based on mathematically
abstract models, and hope they unfold in real life.  If the prediction
comes true, the theoretician revels in the publicity, and makes more predictions
based on the original model.  If it doesn't come true, the theoretician
writes a paper explaining why it didn't, and blames it away as a statistical
manifestation.  Looks like my work for the spring issue of Athletics is
cut out for me either way! \\

\vskip .25 cm
 
\noindent
{\bf Acknowledgements}
 
The IAAF Top 50 lists mentioned herein, as well as the 1997 indoor rankings,
can be found at the IAAF website,
{\tt http://www.iaaf.org/}.  Other results were obtained from the OTFA,
Track and Field News, and
the Track and Field Statistics website ({\tt http://www.uta.fi/$\sim$csmipe/sport/index.html}) maintained by Mika Perkivmdki (csmipe@uta.fi).

\pagebreak

\begin{table}[t]
\begin{center}
{\begin{tabular}{|c l l l l l|}\hline
{\bf Rank} &{\bf Athlete}&{\bf 1997 best} & {\bf 1996 best}& {\bf Difference} & {\bf \% improved} \\ \hline
1 &Maurice Greene & 9.88 &  10.07 &+0.19s & +1.9 \\
& & (9.86, +0.2) & (10.08, -0.3) & & \\ \hline
2 & Donovan Bailey	& 9.89&	 9.88&	+0.01s&	+0.1 \\
&	& (10.03, -2.1)& (9.84, +0.7) & &  \\ \hline
3 &Ato Boldon	& 9.90	& 9.90&	+0.00s&	+0.0 \\
&	&	(9.90, +0.0) &  (9.93, -0.6) & & \\ \hline 
4 & Frank Fredericks & 9.92 & 9.84& -0.08s & -0.8 \\ 
&	& (9.90, +0.2) & (9.86, -0.4) & &  \\ \hline 
5 & Tim Montgomery &  9.94  &  10.07 & +0.13s&+1.3 \\
&	& (9.92, +0.2)& (10.08, -0.3) & &  \\ \hline
6 & Robert Esmie &  9.96 & 10.27A & +0.31s  & +3.0 \\ 
&&  (10.10, -2.1) & (10.18A, +1.5) & &   \\
 &           &    & 10.28 & +0.32s & +3.0 \\ 
 &           &    & (10.28, +0.0) & & \\ \hline
\end{tabular}}
\end{center}
\caption{Men's wind-corrected 1997 World Rankings, with best 1996 performance.}
\label{table1}
\end{table}

\begin{table}
\begin{center}
{\begin{tabular}{|c l c l l l|}\hline
{\bf 1997 rank} &{\bf Athlete}&{\bf 1996 rank} & {\bf 1996 best} & {\bf 1997 best}& {\bf Difference} \\ \hline
1 & Greene	 &    28&10.08 (-0.3)&9.86 (+0.2) &   +0.18 \\
2 & Boldon	 &     3& 9.90 (+0.7)&9.87 (+1.3) &   +0.03 \\
3 & Fredericks&    2& 9.86 (-0.4)&9.90 (+0.2) &   -0.04 \\
4 & Bailey    &    1& 9.84 (+0.7)&9.91 (+0.2) &   -0.07 \\
5 & Jon Drummond&  6& 9.98 (-0.1)&9.92 (+0.8) &   +0.06 \\ \hline
\end{tabular}}
\end{center}
\caption{Men's official 1997 rankings, with change from 1996 rankings}
\label{table2}
\end{table}

\begin{table}
\begin{center}
{\begin{tabular}{|l l l|}\hline
{\bf Athlete} & {\bf 1996 bests} & {\bf 1997 bests} \\ \hline
Bailey & 9.88 (9.84, +0.7) & 9.89 (10.03, -2.1) \\
 & 9.91 (9.93, -0.4) &   9.93 (9.91, +0.2) \\
 & 9.97 (10.00, -0.5) &  9.94 (9.91, +0.5) \\
 & 9.98 (9.98, +0.0) &   9.95 (9.99, -0.7) \\ 
 & 10.04 (9.95, +1.5)  & 9.98 (10.07, -1.5) \\ 
{\it Average} & 9.956 & 9.938 \\ 
{\it Std.\  Dev.} & 0.056 & 0.029 \\ \hline
Fredericks & 9.84 (9.86, -0.4) & 9.92 (9.90, +0.2) \\
 & 9.91 (9.94, -0.5) & 9.94 (9.98, -0.7) \\
 & 9.93 (9.89, +0.7) & 9.96 (9.93, +0.5) \\
 & 9.97 (9.87, +1.9) & 9.97 (9.94, +0.5) \\
 & 10.00 (9.93, +1.1) & 9.97 (9.95, +0.2) \\ 
{\it Average} & 9.930 & 9.952 \\
{\it Std.\ Dev.} & 0.055 & 0.019 \\ \hline
Boldon & 9.90 (9.93, -0.6) & 9.90 (9.90, +0.0) \\
 & 9.92 (9.94, -0.4) & 9.94 (9.89, 0.8) \\
 & 9.94 (9.95, -0.4) & 9.95 (9.87, +1.3) \\
 & 9.94 (9.90, +0.7) & 9.95 (10.00, -0.8) \\
 & 9.97 (9.92, +0.8) & 9.99 (9.95, +0.6) \\
{\it Average} & 9.934 & 9.946 \\ 
{\it Std.\ Dev.} & 0.023 & 0.029 \\ \hline
\end{tabular}}
\end{center}
\caption{Top 5 1996 and 1997 wind-corrected performances for Bailey, Fredericks,
and Boldon, listing average and standard deviation.}
\label{table2a}
\end{table}
 
\begin{table}
\begin{center}
{\begin{tabular}{|l l l||l l l|}\hline
\multicolumn{3}{|c||} {\bf Maurice Greene}  & \multicolumn{3}{c|} {\bf Marion
Jones} \\ \hline
10.19 (10.19, +0.0) & 25 May & 1 (f) & & & \\ \hline
10.05 (9.96, +1.5) & 12 Jun & 1 (h) &  11.05 (10.98, +0.9) & 12 Jun & 1 (h)\\ 
10.10 (10.08, +0.2) & 12 Jun & 1 (sf) &10.93 (10.92, +0.1) & 12 Jun & 1 (sf) \\
9.92 (9.90, +0.2) & 13 Jun & 1 (f) &  11.05 (10.97, +1.1) & 13 Jun & 1 (f) \\ \hline
10.13 (10.23, -1.5) & 25 Jun & 4 (f) & 11.07 (11.23, -1.9) & 27 Jun & 1 (f)\\
9.96 (9.90, +1.0) & 02 Jul & 1 (f) &  10.94 (10.90, +0.5) & 02 Jul & 2 (f) \\
10.05 (10.01, +0.6) & 07 Jul & 2 (f)&11.06 (11.06, +0.0) & 04 Jul & 1 (f)\\ 
10.04 (10.04, +0.0) & 13 Jul & 2 (f)& & & \\ \hline
10.38 (10.32, +1.0) & 02 Aug & 1 (h)& 11.04 (11.09, -0.7) & 02 Aug & 1 (h) \\
9.98 (9.90, +1.3) & 02 Aug & 1 (qf) &  10.99 (10.96, +0.3) & 02 Aug & 1 (qf)\\
9.93 (9.90, +0.5) & 03 Aug & 1 (sf) & 10.94 (10.94, -0.1) & 03 Aug & 1 (sf)\\
9.88 (9.86, +0.2) & 03 Aug & 1 (f)& 10.86 (10.83, +0.4) & 03 Aug & 1 (f)\\ \hline
9.95 (9.99, -0.7) & 13 Aug & 2 (f)& 10.88 (10.97, -1.2) & 13 Aug & 2 (f)\\
10.09 (10.06, +0.4) & 16 Aug & 3 (f) & & & \\
9.94 (9.91, +0.1) & 22 Aug & 2 (f) & 10.82 (10.76, +0.9) & 22 Aug & 1 (f)\\
 & & & 10.87 (10.81, +0.8) & 16 Aug & 1 (f)\\ \hline
\end{tabular}}
\end{center}
\caption{Progression for 1997 100m World Champions, indicating placing in
race (h - heat; qf - quarter-final; sf - semi-final; f - final).}
\label{table2b}
\end{table}

\begin{table}
\begin{center}
{\begin{tabular}{|c l l l l l|}\hline
{\bf Rank} &{\bf Athlete}&{\bf 1997 best} & {\bf 1996 best}& {\bf Difference} &
{\bf \% improved} \\ \hline
1 & Merlene Ottey  & 10.79 &  10.83&	+0.04s & +0.4 \\
& &(10.81, -0.4) &(10.74, +1.3) & & \\ \hline
2 & Marion Jones & 10.82 & ---- & ---- & ---- \\ 
& &(10.76, +0.9) & --- &  &  \\ \hline
3 & Zhanna Pintussevich& 10.88  & ---- & ---- & ---- \\ 
& & (10.85, +0.4) & ---- & &  \\ \hline
4 & Gail Devers	& 10.92& 10.92&	+0.00s & +0.0 \\
& & (10.89, +0.5) &(10.83, +1.3)  & & \\ \hline
5 &Sevatheda Fynes  & 10.98 &11.16  & +0.18s & +1.6   \\ 
& & (11.07, -1.2) & (11.24, -1.0)  & & \\ \hline
\end{tabular}}
\end{center}
\caption{Women's wind-corrected 1997 World Rankings, with best 1996 performance and improvement.}
\label{table3}
\end{table}

\begin{table}
\begin{center}
{\begin{tabular}{|c l c l l l|}\hline
{\bf 1997 rank}& {\bf Athlete}&{\bf 1996 rank} & {\bf 1996 best} & {\bf 1997 best}& {\bf Difference} \\ \hline
1& Jones    &    NR &   ---  & 10.76 (+0.9)&---- \\ 
2& Ottey   &     1 &   10.74 (+1.3)&  10.83 (+0.4)&-0.09s \\
3& Pintussevich&  NR &   ----& 10.85 (+0.8)&---- \\
4& Devers      &    3 &   10.83 (+1.3)&	10.88 (+0.8)&-0.05s \\
5& Christine Arron  &   NR &   ---& 11.03 (-0.3)&---- \\ \hline
\end{tabular}}
\end{center}
\caption{Women's official 1997 rankings, with change from 1996 rankings.}
\label{table4}
\end{table}

\begin{table}
\begin{center}
{\begin{tabular}{|l l l l|}\hline
&{\bf 50m split} & {\bf 60m split} & {\bf 1997 indoor 60m best} \\ \hline
\multicolumn{4}{|l|} {\bf Men} \\ 
Greene&	5.56  (5.55)&	6.41 (6.40)&6.54 (Atlanta, 01 Mar) \\
Bailey&	5.59  (5.58)&	6.44 (6.43)&6.51 (Maebashi, 08 Feb) \\
Montgomery&5.57  (5.56)	&6.43 (6.42)&6.51 (Atlanta, 01 Mar) \\
Fredericks&5.59  (5.58)	&6.45 (6.44)&---- \\
Boldon	&5.59  (5/58)	&6.46 (6.45)&6.49 (Birmingham, 23 Feb) \\ \hline
\multicolumn{4}{|l|} {\bf Women} \\
Jones	&6.04 (6.02)&	6.99 (6.96)&---- \\
Pintussevich&6.04 (6.02)  &	6.99 (6.96)&    7.21 (Madrid, 05 Feb) \\
Arron	&6.19 (6.17)  &	7.14 (7.12)&---- \\
Miller	&6.19 (6.17)  &	7.16 (7.14) &   ---- \\
Paschke	&6.20 (6,18)  &	7.18 (7.16)&7.17 (Sindelfingen, 01 Mar) \\
Ottey	&6.26 (6.24)  &	7.23 (7.21)&7.10 (Birmingham, 23 Feb) \\ \hline
\end{tabular}}
\end{center}
\caption{Comparison of wind-corrected (official) 50m and 60m splits from 1997 
WC final with athlete's 1997 indoor best at 60m.}
\label{table5}
\end{table}

\begin{table}
\begin{center}
{\begin{tabular}{|l l l|}\hline
\multicolumn{3}{|l|} {\bf 50m (WR 5.56A Donovan Bailey CAN Reno, 9 Feb 1996)} \\
 5.60 & Michael Green            JAM  &Liivin (16 Feb) \\ 
 5.65 & Deji Aliu             NGR  & Madrid   (05 Feb) \\ 
 5.67 & Bruny Surin            CAN &  Hamilton (18 Jan) \\ 
 5.68 & Darren Braithwaite      GBR&  Madrid  (05 Feb) \\ 
 5.70 & Robert Esmie            CAN&  Hamilton (18 Jan) \\ \hline
\multicolumn{3}{|l|} {\bf 60m (WR 6.41 Andre Cason USA Madrid, 14 Feb 1992)} \\
 6.49 & Michael        Green        JAM& Liivin   (16 Feb)  \\
 6.49 & Ato            Boldon       TRI& Birmingham (23 Feb)\\
 6.49 & Randall        Evans        USA& Atlanta    (01 Mar)\\
 6.50 & Bruny          Surin        CAN& Paris Bercy (07 Mar)\\
 6.50 & Charalambos    Papadias     GRE& Paris Bercy (07 Mar)\\
 6.51 & Donovan        Bailey       CAN& Maebashi  (08 Feb) \\
 6.51 & Tim            Montgomery   USA& Atlanta (01 Mar)   \\
 6.51 & Linford        Christie     GBR& Sindelfingen (01 Mar) \\
 6.52 & Ray            Stewart      JAM& Liivin   (16 Feb)    \\
 6.52 & Davidson       Ezinwa       NGR& Paris Bercy (07 Mar)\\
 6.54 & Maurice     Greene        USA&  Atlanta (01 Mar)    \\
 6.54 & Robert      Esmie         CAN&  Paris Bercy (07 Mar)\\ \hline
\end{tabular}}
\end{center}
\caption{Men's 1997 indoor rankings, 50m and 60m (best performance only).}
\label{table6}
\end{table}

\begin{table}
\begin{center}
{\begin{tabular}{|l l l|}\hline
\multicolumn{3}{|l|} {\bf 50m (WR 5.96 Irina Privalova RUS Madrid, 9 Feb 1995) }\\
 6.05& Irina          Privalova        RUS & Liivin   (16 Feb) \\
 6.08& Christy        Opara            NGR   &Liivin   (16 Feb) \\
 6.14& Fridirique     Bangui           FRA & Liivin   (16 Feb) \\
 6.19& Odiah          Sidibi           FRA & Liivin   (16 Feb) \\
 6.24& Zhanna         Pintussevich      UKR &Madrid   (05 Feb) \\ \hline
\multicolumn{3}{|l|} {\bf 60m (WR 6.92 Irina Privalova RUS Madrid, 11 Feb 1993 / 9 Feb 1995)} \\
 7.00& Gail         Devers          USA& Atlanta (01 Mar) \\
 7.02& Irina        Privalova       RUS& Gent    (12 Feb) \\
 7.02& Christy      Opara           NGR& Gent    (12 Feb) \\
 7.05& Chioma       Ajunwa          NGR& Erfurt  (05 Feb) \\
 7.12& Gwen         Torrence        USA& Fairfax  (22 Feb) \\
 7.14& Carlette     Guidry-White    USA& New York (07 Feb) \\
 7.15& Ekaterini    Thanou          GRE& Pireus   (14 Feb) \\
 7.16&  Odiah      Sidibi        FRA&Liivin       (16 Feb)   \\
 7.17&  Endurance  Ojokolo       NGR&Birmingham   (23 Feb) \\
 7.17&  Melanie    Paschke       GER&Sindelfingen (01 Mar)   \\
 7.19&  Michelle   Freeman       JAM&Chemnitz   (29 Jan)    \\
 7.20&  Merlene    Frazer        JAM&Dortmund   (09 Feb)   \\
 7.20&  Juliet     Cuthbert      JAM&Dortmund   (09 Feb)  \\
 7.20&  Ekaterini  Koffa         GRE&Pireus     (14 Feb) \\
 7.21&  Zhanna     Pintussevich   UKR&Madrid     (05 Feb)    \\
 7.21&  Nadezhda   Roshchupkina  RUS&Volgograd  (21 Feb)   \\ \hline
\end{tabular}}
\end{center}
\caption{Women's 1997 indoor rankings, 50m and 60m (best performance only).}
\label{table7}
\end{table}

\end{document}